\documentclass{article}

\usepackage{amssymb}
\usepackage{graphicx}

\title{Spherical black holes with regular center\\
{\small{}---\hspace{1mm}\emph{a review of existing models including a recent realization with Gaussian sources}\hspace{1mm}---}}

\author{Stefano Ansoldi\footnote{Email: \texttt{ansoldi@trieste.infn.it} --- Webpage: \texttt{http://www-dft.ts.infn.it/$\sim$ansoldi}}\\
\emph{\normalsize{}Department of Physics, Kyoto University, Kyoto, Japan}\\
\emph{\normalsize{}International Center for Relativistic Astrophysics, Pescara, Italy}\\
\emph{\normalsize{}Istituto Nazionale di Fisica Nucleare, Sezione di Trieste, Italy}}

\date{\hrule\vskip 2 mm%
\footnotesize{}To appear in the proceedings of\\
\emph{BH2, Dynamics and Thermodynamics of Blackholes and Naked Singularities},\\
May 10-12 2007, Milano, Italy\\
conference website: \texttt{http://www.mate.polimi.it/bh2}\\[2mm]
\hrule\vskip 2 mm%
Preprint: KUNS-2108}

\begin{document}

\maketitle

\begin{abstract}
We review, in a historical perspective, some results about black hole spacetimes with a regular
center. We then see how their properties are realized in a specific solution that recently
appeared; in particular we analyze in detail the (necessary) violation of the strong energy condition.
\end{abstract}

\section{Introduction}

Recently there has been a renewed interest for the search of solutions of Einstein equations,
mostly motivated by the study of higher dimensional gravity, for instance in the context
of the brane world scenario \cite{bib:PhReL1999..83..3370S,bib:PhReL1999..83..4690S}
and of string theory \cite{bib:YCS062006..........S}. Some beautiful
examples of higher dimensional solutions have appeared (see, e.g.,
\cite{bib:ClQuG2006..23..6919T,bib:ClQuG2007..24..3141T}), which feature interesting properties
that, in the pure $4$-dimensional framework, are absent. This renewed interest is also having
some influence on a rather more specialized, but very interesting, research area, that of regular
black holes.
It is of course well-known that, under rather generic conditions on the energy--matter content
of spacetime, classical solution of Einstein equations exhibit both, future
\cite{bib:PhReL1965..14....57P} and past \cite{bib:PhReL1965..15...689H,bib:PRSLA1966.294...511H,%
bib:PRSLA1966.295...490H,bib:PhReL1966..17...445G,bib:PRSLA1967.300...187H} singularities
\cite{bib:PRSLA1970.314...529P,bib:CMaPh1975..41....65C,bib:PhReD1994..50..3692B,bib:GRGra1997..29...701S,%
bib:Freem1970...1..1279W,bib:UofCP1984...1...491W,bib:CaUPr1973...1...391E}
usually hidden by an event horizon \cite{bib:CMaPh1968...8...245I}.
This fact, which is clearly exemplified by the first and probably most well-known solution of Einstein
equations, i.e. the Schwarzschild solution
\cite{bib:SKPAW1916.......189S,bib:SKPAW1916.......424S},
has been likely fully appreciated only after the study of its
global analytic extension \cite{bib:Natur1924.113...192E,bib:PhRev1958.110...965F,%
bib:PhRev1960.110..1743K,bib:PuMaD1960...7...285S,bib:ShAsI1963..........N,%
bib:Freem1970...1..1279W,bib:CaUPr1973...1...391E},
which gives a consistent picture
of its properties, like the existence of the region inside the horizon,
in which the radial and time coordinate exchange their character, and the presence of the
central singularity. Although the presence of, both, the black hole region and the central
singularity has been eventually accepted, i.e. we learned how to classically live with them,
especially the presence of a singularity is a recurrent motivation to underline the inadequacy
of general relativity as a theory of spacetime below some length scale (apart from the theoretical
motivations, it is \emph{a fact} that, experimentally, gravity can be tested only in
a finite range of scales). On one side, maybe the most well known one, this has
motivated the search for a more complete theory of gravity including also quantum effects
(see, e.g., \cite{bib:ClaPr2004...1...308K} for a recent, comprehensive review).
On the other side, it has also sustained many efforts to push as much as possible Einstein gravity
to its limit, trying to avoid, if not the black hole region, at least the central singularity
in a way as consistent as possible with physical requirements\footnote{The two points of view just
outlined should be seen as complementary and, often, even integrated.}. Following some very early
ideas that date back to the work of Sakharov \cite{bib:SPJET1966..22...241S},
Gliner \cite{bib:SPJET1966..22...378G}
and Bardeen \cite{bib:GR5Tp1968.......174B}, solutions having a global
structure very similar to the one of black hole spacetimes, but in which the central singularity
is absent, have been found (references to them \emph{will appear} in the rest of the paper).

In this contribution we are going to briefly review some of these
ideas, but, before concluding this introductory section with the layout of the rest of the paper,
we would like to make a couple of remarks. The first of them is that, in contrast to the fact that
nowadays the world of theoretical physics witnesses a consistent number of strongly believed
(but as yet unproven) conjectures, most of the results about black holes and their properties are,
in fact \emph{theorems}. Theorems usually make some hypotheses, which in this case can be roughly
interpreted within a threefold scheme; i) the validity of some geometric properties, usually related
to the behavior of geodesics (for instance the existence of trapped surfaces, and so on);
ii) the validity of some conditions on the matter fields that are coupled to gravity (energy conditions);
iii) the validity of some (more technical) hypotheses about the global/causal structure of spacetime.
It is then clear that the possibility of singularity avoidance, within the context defined by
general relativity, requires the violation of, at least, one of the above conditions.
Since conditions of type iii) are mostly technical,
there has been a great effort to make them as general as possible (although sometimes this means less
immediate) if not to remove them at all, by generalizing the earliest results \cite{bib:PRSLA1970.314...529P,%
bib:CaUPr1973...1...391E,bib:UofCP1984...1...491W}. Conditions i) are usually
the requirement that some \emph{indicator} exists, which emphasizes that something a little bit unusual for
a ``flat'' non covariant mind is taking place in spacetime, and are usually related to the existence of
horizons, so there is little reason to modify them. It is then natural that, as a possible way to avoid singularities,
a relaxation of conditions of type ii) has been advocated. With a strongly conservative attitude, a word
of caution should be sounded at this point. It is, in fact, known that matter and energy violating some
of the energy conditions, have \emph{as yet} unobserved properties\footnote{We will come back to this point
later on, mentioning vacuum and the cosmological constant. A clear discussion of this point can be found
in the standard reference \cite{bib:CaUPr1973...1...391E}; see also the early \cite{bib:PhReD1983..28..1265B}
for a physically oriented discussion of the implications of a violation of the weak energy condition.}:
this means that we are \emph{not yet} able to produce them in a laboratory by a well-known, generally
reproducible procedure. We have, nevertheless, good candidates to realize these violations when we treat at an
effective level the quantum properties of spacetime and matter at some length/energy scales:
this is very suggestive, since it directly connects to the, possibly, incomplete character of
classical general relativity as a theory of spacetime and with the ongoing, diversified, efforts
toward its quantization \cite{bib:ClaPr2004...1...308K}. To review in more detail some aspects related
to the above reflections, we plan as follows.

In section \ref{sec:rev} we review various regular models of spacetime,
centering our attention, almost exclusively, on regular black holes of a very specific type (specified below).
After a review of the earliest ideas (subsection \ref{subsec:earide}) we analyze their first concrete
(and, perhaps, to some extent independent) realization, known as the Bardeen solution: we review also some
studies, which appeared much later, discussing its global character (subsection \ref{subsec:barsol}); we
then continue with a discussion of black hole interiors (subsection \ref{subsec:blaholint}) reporting various early
proposals, which adopted spacetime junctions to get rid of singularities; this brings
 us to the central part of
this section (subsection \ref{subsec:exasol}), where some exact solutions are analyzed, together with the
possibility of physical realizations for the energy-matter content which should act as their source
(subsubsection \ref{subsubsec:matcon}). The solutions that we will have described up to this point are not
extemporary realizations, but can be understood in a very interesting, complete and general framework: we
thus review the essence of this framework in subsection \ref{subsec:genfra}. This section is concluded
with a very concise summary of the results that we have reviewed (subsection \ref{subsec:genanasyn}).

Then, in section \ref{sec:gausou} we use a recently obtained solution, which is another
possible realization of the general type of solutions described in subsection \ref{subsec:genfra},
to perform a simple exercise,  i.e. the study of the violation of one of the energy conditions. For completeness,
after introducing the algebraic form of the solution, we quickly construct its global spacetime structure
in subsection \ref{subsec:maxext} (this result follows immediately from the results reviewed in subsection
\ref{subsec:genfra}); we then show which regions of spacetime are filled with matter violating the
strong energy condition (subsection \ref{subsec:enecon}). The results of this second part of the paper
are summarized in subsection \ref{subsec:eneconsum}.

Some general comments and remarks find space in the concise concluding section, i.e. section \ref{sec:discon}.
We now conclude this introduction by fixing one notation and one naming convention, as below.

\subsection{\label{sec:pre}Conventions and notations}

In what follows we will concentrate on spherically symmetric solutions of Einstein equations and restrict
ourself to media which satisfy the condition that the radial pressure equals the opposite of the energy
density. We will then use, throughout and unless otherwise stated, the coordinate system $(t,r,\vartheta,\varphi)$, in which
the metric can be written in the static form adapted to the spherical symmetry, i.e.
\begin{equation}
    g _{\mu \nu}
    =
    {\mathrm{diag}}
    \left( - f (r) , f (r) ^{-1} , r ^{2} , r ^{2} \sin ^{2} \vartheta \right)
    .
\label{eq:metfor}
\end{equation}
As apparent from the above definition we adopt the signature $(-,+,+,+)$. We occasionally
will use the name \emph{metric function} for the function $f (r)$. We do not spend extra
comments about the meaning of the coordinate choice, which is standard and discussed in
detail in various textbooks (see for instance \cite{bib:Freem1970...1..1279W}; any
other textbook choice will be equivalent); Thus, without restating every time our coordinate choice,
in what follows we will specify various metrics just by specifying the corresponding metric function.
In view of the above, when we will have to discuss the maximal extension of solutions that
admit an expression of the metric in the form (\ref{eq:metfor}), although we will follow the naming conventions
of the standard reference \cite{bib:CaUPr1973...1...391E} for boundaries
as infinity, only in one point of our discussion we will need a few more global ideas than the one concisely
and effectively discussed in \cite{bib:JMaPh1970..11..2280W}. We will moreover use the standard notation
$T _{\mu \nu}$ for the stress-energy tensor which appears on the righthand side of Einstein equations.

\section{\label{sec:rev}Regular spacetime models}

As we briefly discussed in the introductory section, there are various reasons to try to find
consistent solutions of Einstein equations that
describe regular spacetimes. As we also briefly discussed, the possibilities are somewhat restricted
by the existence of various singularity theorems \cite{bib:PhReL1965..14....57P,bib:PhReL1965..15...689H,
bib:PRSLA1966.294...511H,bib:PRSLA1966.295...490H,bib:PhReL1966..17...445G,bib:PRSLA1967.300...187H,%
bib:PRSLA1970.314...529P,bib:CMaPh1975..41....65C,bib:PhReD1994..50..3692B,bib:GRGra1997..29...701S,%
bib:Freem1970...1..1279W,bib:CaUPr1973...1...391E,bib:UofCP1984...1...491W}, which apply, both,
to cosmology as well as to gravitational collapse. Thus, the real crucial point is if/how it is possible
to violate some of the hypothesis of singularity theorems, without obtaining unphysical models.
The literature developed a lot in this direction, in the field of, both again, cosmology and
gravitational collapse. Solutions that do not posses horizons are, of course, a natural
framework in which singularity theorems can not be applied and in which regular solutions
can be searched: boson stars are an example (see \cite{bib:PhRev1968.172..1331K,bib:PhRev1969.187..1767B}
as well as, e.g., \cite{bib:NuPhy2000B564...185S,bib:ClQuG2004..21..1135W} for recent reviews).
On the other hand it seems natural that in various physically reasonable
situations horizons are indeed formed\footnote{The
problem of the nature of the horizon and of the inside horizon region is a fascinating one,
which has been carefully analyzed in the past (see, for instance, \cite{bib:CMaPh1968...8...245I,%
bib:PhReL1989..63..1663I,bib:PhReD1990..41..1796I,bib:PhReL1991..67...789O,%
bib:PhReD1994..50..7372M,bib:ClQuG1998...5..L201I} and references therein) also in
presence of a non-abelian hair \cite{bib:PhReD1997..56..3459Z},
i.e. for the so called Einstein-Yang-Mills black holes (additional references can
be found in the bibliography of the already cited \cite{bib:PhReD1997..56..3459Z}).}.
It becomes, thus, interesting to consider if singularities can be avoided in presence of
horizons. Solutions for which this is true are generically called \emph{regular black hole}. A great
variety of these solutions has been investigated (a paper presenting a broad
perspective on the subject is \cite{bib:PhReD2002..65124024H}) and a very effective classification
of them can be found in \cite{bib:GRGra2007..39...973M}: we are going to follow this
classification scheme and concentrate our attention on existing models of \emph{black holes with
a regular center}, i.e. the type 1 regular black holes in the above mentioned classification
(see \cite{bib:GRGra2007..39...973M}, page 975).
For reasons of space we will also remain in the more standard arena in which only the
gravitational field and the electromagnetic field (including non-linear electrodynamics)
are present\footnote{Nevertheless, we would like to stress that studies of more general situations,
for instance in presence of Yang-Mills fields \cite{bib:CMaPh1993.151...303W,bib:CMaPh1994.163...141M} and of
massive \cite{bib:IJMPA1999..14..2013A}, nonlinear \cite{bib:PhReD2001..64064013B,bib:ClQuG2001..18..1715L}
and phantom \cite{bib:PhReL2006..96251101F} scalar fields do exist. Generalizations
(see for instance \cite{bib:PhLeA1989.142...341T} and \cite{bib:IJMPD2004..13..1095D}) of existing
models (the ones in \cite{bib:LNuCi1981..32...161G} and \cite{bib:GRGra1992..24...235D}, respectively)
witness an interest for higher dimensional extensions, which nowadays can be motivated by various
studies, e.g. those about brane-world black holes \cite{bib:PTPSu2002.148...307T,%
bib:PhReD2003..68024025M,bib:PhReD2003..68024035N,bib:gr-qc2007..09..3674T}.}.

\subsection{\label{subsec:earide}The earliest ideas}

In the above defined context, the first seeds that will eventually lead to the idea of regular
black hole solutions go back to the mid sixties with Sakharov proposal to consider
the equation of state $\mathrm{(pressure)} = - \mathrm{(energy\ density)}$ as the appropriate
equation of state for matter and energy at very high densities \cite{bib:SPJET1966..22...241S}.
This equation of state is the same
equation of state obeyed by the \emph{cosmological vacuum}, by which we are going to identify
the cosmological constant term of Einstein equation, when we decide to interpret it as a
contribution to the right hand side of Einstein equations (i.e. to the energy momentum tensor),
rather than as a, left hand side, geometric entity. It can be
agreed to a great extent that, as very briefly, but precisely,
noted in \cite{bib:ClQuG2002..19...725D} (footnote on page 738), this may seem a rather
philosophical distinction between \emph{a vacuum spacetime} and a \emph{spacetime filled
with vacuum}; on the other hand, we would like to stress a related physical point of view
i.e. the fact that the first one, a \emph{a vacuum spacetime} where the cosmological term is a
geometrical parameter, does satisfy all the known energy conditions, whereas the second
one, i.e. the one \emph{filled with vacuum}, does not satisfy the strong energy condition.
Thus, in the second case, the related question of the physical origin of the vacuum,
i.e. of a medium having a negative pressure equal in magnitude to its energy density
(we are, of course, making the safe assumption $\rho > 0$),
becomes a relevant one. We will, briefly, come back to this point later on.
Nevertheless, it is clearly very interesting to investigate the consequences of
accepting (physically and not only philosophically) the idea of a \emph{spacetime filled
with vacuum}, hence the early proposal of Gliner \cite{bib:SPJET1966..22...378G}
that a \emph{spacetime filled with vacuum} could provide a proper description of the
final stage of gravitational collapse, replacing the future singularity\footnote{The
term used by Gliner, which corresponds to our \emph{spacetime filled with vacuum}, is $\mu$-vacuum
\cite{bib:SPJET1966..22...378G}.}. This proposal,
which should have helped getting rid of the singularity in favor of a regular
spacetime, was not immediately realized by a specific solution of Einstein equations.
In general a solution of this kind would require a consistent treatment of, both, the
gravitational fields and the matter fields \emph{at the quantum level}. We are still waiting
for a framework capable to provide us with this treatment; nevertheless already the
classical analysis turns out to be quite interesting, as shown
for instance by the seminal paper \cite{bib:AnnPh1979.118....84S}.

\subsection{\label{subsec:barsol}The Bardeen solution}

\begin{figure}
\begin{center}
\fbox{\vbox{\hbox{%
\hbox{\fbox{\includegraphics[width=6.8cm]{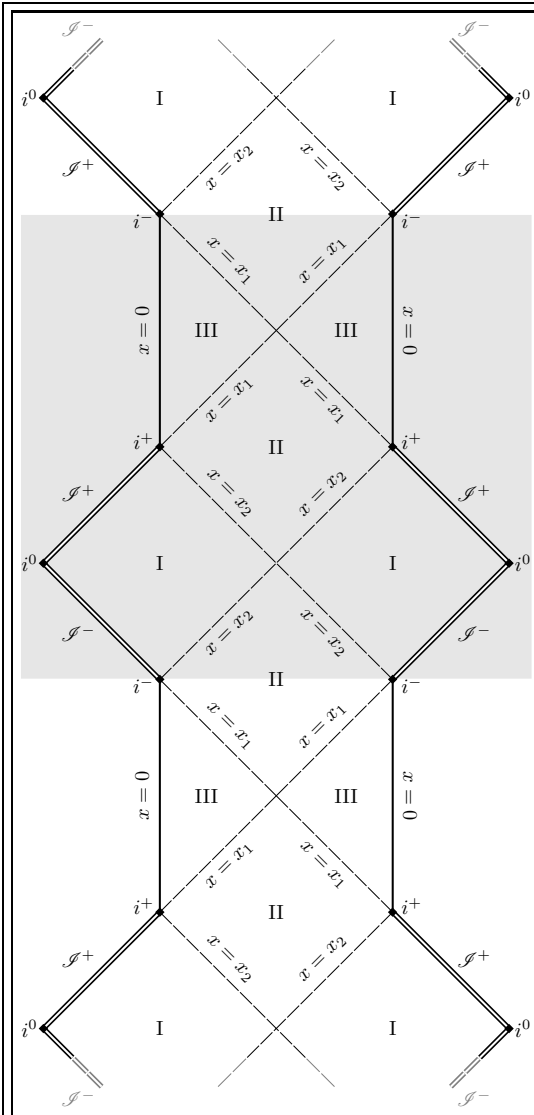}}\hskip 1 mm%
\hbox{\parbox[b]{4.7cm}{\caption{\label{fig:pendiamuGmuc}{\small{}maximal
extension of the metric of a black hole with a regular center when, both, the event
and the Cauchy horizons are present. Variable $x$ is a re\-sca\-led, dimensionless radial
coordinate, obtained rescaling $r$ by a characteristic length scale of the
solution. Labels of the boundaries of spacetime follow
the notation in {\protect\cite{bib:CaUPr1973...1...391E}}.
The maximal extension is Resi\ss{}ner-Nordstr\o{}m--like, but $x = 0$ is
non-singular. Each point in the diagram corresponds to a 2-sphere ${\mathbb{S}} _{2}$.
The double lines represent future and past null infinity, whereas
the dashed ones correspond to values of $x$ at which the metric function vanishes,
i.e. they are the event horizon ($x = x _{2}$) and the Cauchy horizon ($x = x _{1}$)
Light rays lines form a $45$ degree angle with the horizontal direction.
From the diagram it is readily seen that the spacetime is not globally hyperbolic
and consists of an infinite sequence of regions, some in which the metric function
is positive and some in which the metric function is negative (in particular the regions bounded
by $x = 0$ and those bounded by null infinity are of the first kind, whereas the ones
}}}}}}
\hbox{\parbox[b]{11.85cm}{{\small{}bounded by $x = x _{1}$ and $x = x _{2}$ are of the second kind). As discussed in the main
text, this maximal extension is a completely general results for metrics which satisfy the conditions discussed on page
{\protect\pageref{top:theore}} in subsection {\protect\ref{subsec:genfra}, when the
parameters of the model fall in some specific range (for the metric in ({\protect\ref{eq:adimetfor}}) and
({\protect\ref{eq:adimetfun}}) this range is $\mu _{0} > \mu _{0} ^{\mathrm{cr}}$)}. The way in which this solution avoids
the emergence of singularities is related to the structure of the regions between the two
\emph{antipodal} $x = 0$ (i.e. $r = 0$) lines: here future directed causal curves can \emph{wrap around
the universe} until they finally free themselves crossing a type II region and ending up
in an asymptotically flat type III region (see also the main text, starting
at page {\protect\pageref{top:barsolsinavo}} in subsection {\protect\ref{subsec:barsol}}).}}}}}
\end{center}
\end{figure}

This said, it is then interesting to note that, just a few years
later the appearance of these seminal ideas
and likely with different motivations, the first regular solution
of Einstein equations having an event horizon was obtained by J. M. Bardeen \cite{bib:GR5Tp1968.......174B}.
Although this was not realized for some time, this solution is a concrete
example of a completely general class of models (discussed below)
that realize singularity avoidance using
a vacuum-like equation of state for matter below some length scale.
Bardeen's solution (see also \cite{bib:PhReD1994..50..3692B}) is a solution
of Einstein equations in the presence of an electromagnetic field and, as the
well-known Rei\ss{}ner-Nordstr\o{}m solution \cite{bib:AnPhG1916..50...106R,bib:PKNAW1918..20..1238N},
is parametrized by mass
$m$ and charge $e$. Also the line element can be put in exactly the same form,
i.e the static and spherically symmetric one defined in (\ref{eq:metfor}),
with the metric function $f(r)$ being\label{top:BarSol}
\[
    f (r) = 1 - \frac{2 m r ^{2}}{(r ^{2} + e ^{2}) ^{3/2}}
          = 1 - \left( \frac{m}{e} \right) \frac{2  ( r / e ) ^{2}}{(( r / e ) ^{2} + 1) ^{3/2}}
\]
and $r \geq 0$ (notice the equal sign in the previous relation). It is not
difficult to see that i) there are values of the $m/e$ ratio for which $f (r)$
has no zeroes (and is thus always positive), ii) values for which it has two zeroes,
as well as an in between case, where iii) the function is always non-negative but
vanishes at only one point together with its first derivative. Notice that
this function is everywhere well defined as are the curvature tensors and scalar of
the solution that it represents; the metric described by $f (r)$ is asymptotically
flat, for small $r$ it behaves as the de Sitter metric, since
\[
    f (r) \approx 1 - 2 \frac{m}{e} r ^{2}, \quad r \approx 0 ^{+}
    ,
\]
whereas for large $r$ it asymptotically behaves as the Schwarzschild metric. In the
case ii) mentioned above the black hole interior
does not terminate on a singularity but crosses an interior Cauchy horizon
and develops in a region that becomes more and more de Sitter like, eventually ending
with a regular origin at $r = 0$. This solution represents a strong constraint on
possible generalizations of existing singularity theorems, since it acts as a counterexample
that prevents relaxing too much the necessary conditions for the existence of a
singularity. It, also, has very interesting properties and its causal structure and
its geometrical structure have been studied in detail
\cite{bib:PhReD1994..50..3692B,bib:PhReD1997..55..7615B},
although many years after its appearance. From these studies, it turned out that there
can be a very generic mechanism for singularity avoidance in classical general relativity,
which requires what has been called a \emph{topology change}. If we, again, restrict our attention
to the values of the parameters for which the Bardeen solution has both an event and
a Cauchy horizon, it can be easily seen that its global structure is as in figure
\ref{fig:pendiamuGmuc}, i.e. it resembles the maximal extension \cite{bib:PhRev1960.120..1507B,bib:PLett1966..21...423C}
of Rei\ss{}ner-Nordstr\o{}m \cite{bib:AnPhG1916..50...106R,bib:PKNAW1918..20..1238N} spacetime but with
a regular origin.
\label{top:barsolsinavo}The analysis of Borde \cite{bib:PhReD1994..50..3692B,bib:PhReD1997..55..7615B},
then, shows that the singularity avoidance can be related to the fact that in regions where
spacelike slices of the Rei\ss{}ner-Nordstr\o{}m solution would have the topology of
$\mathbb{R} \times \mathbb{S} ^{1}$, the same spacelike slices in the Bardeen solution have the
topology $\mathbb{S} ^{3}$. Since there are corresponding regions of the extended manifold where
spacelike slices of both, the Rei\ss{}ner-Nordstr\o{}m and the Bardeen solution have, indeed, the topology
$\mathbb{R} \times \mathbb{S} ^{1}$, in \cite{bib:PhReD1994..50..3692B,bib:PhReD1997..55..7615B}
the avoidance of singularity has been associated to the \emph{topology change} which appears
in the structure of the spacelike slices, which from open become closed; this allows singularity
avoidance\footnote{This provides an earlier answer to the much later question recently raised in
\cite{bib:gr-qc2007..08..2360K}.} since in the closed case
\begin{quote}
it is possible for light rays to ``wrap around the Universe;'' i.e., although both the
``ingoing'' and ``outgoing'' systems of future-directed null geodesics [\emph{starting}$^{*}$] from
[\emph{a trapped surface}$^{*}$] ${\mathcal{T}}$ are converging, the two sets converge to focii at
different $r = 0$ points (antipodally located with respect to each other)
[\emph{quotation taken from} \cite{bib:PhReD1997..55..7615B};
${}^{*} \emph{present author addition}$].
\end{quote}

\subsection{\label{subsec:blaholint}Black hole interiors}

The Bardeen solution represents a first concrete case that realizes the early physical
idea of Sakharov \cite{bib:SPJET1966..22...241S} and Gliner \cite{bib:SPJET1966..22...378G},
replacing the singularity by a regular de Sitter core.
In view of the early appearance of the Schwarzschild \cite{bib:SKPAW1916.......189S,bib:SKPAW1916.......424S}
and de Sitter \cite{bib:PKNAW1917..19..1217S,bib:PKNAW1917..20...229S} solutions and
of the global structure of these spacetimes, it is natural that the
idea to replace the black hole interior of Schwarzschild spacetime with the interior
(i.e., before the cosmologica horizon) region of de Sitter spacetime appeared quite early
\cite{bib:LNuCi1981..32...161G} (with related thermodynamical interpretations
\cite{bib:LNuCi1982..33...127G}) together with charged generalizations
\cite{bib:NuCim1985.B85...142Z,bib:GRGra1985..17...739Z}; this intuitive ideas
can be supported, in fact, by the useful formalism of Israel junction conditions
\cite{bib:NuCim1966.B44.....1I,bib:NuCim1967.B48...463I,bib:PhReD1991..43..1129I}.
Despite the fact that the simpler idea to perform the junction at a null surface \cite{bib:PhLeA1988.126...229Z}
fails due to stability issues \cite{bib:PhLeA1989.138....89G} and the appearance of a discontinuity
in the pressure at the null junction \cite{bib:LNuCi1985..44...177G,bib:ClQuG1998...5..L201I},
it is, instead, possible to substitute \emph{part} of the black hole interior
with part of de Sitter spacetime interposing a layer of non-inflationary material
\cite{bib:LNuCi1985..44...177G}. It is particularly interesting the case in which
this layer is spacelike \cite{bib:PhLeB1989.216...272M,bib:PhReD1990..41...383M},
which gives the so called Schwarzschild--de Sitter model\footnote{Not to be confused
with the Schwarzschild--de Sitter solution of Einstein equations.}. A motivation for these
studies is naturally the fact that, approaching a singularity, the curvature becomes
unbounded. On the other hand and quite generically, a consistent framework for singularity
avoidance (due, e.g., to quantum \cite{bib:PhLeB1981.106...307V,bib:PhReD1991..43..3144M}
or, nowadays more fashionable and exotic, quintessential \cite{bib:PhReD2005..72024016K} effects
or motivated by a theory more fundamental than general relativity \cite{bib:ClaPr2004...1...308K})
should prevent this divergence\footnote{Similar ideas can, of course,
be applied to the initial singularity in cosmology and we will touch briefly this point later on;
meanwhile, we would like to remember some interesting proposals
\cite{bib:PhLeB1980..91....99S,bib:PhReL1992..68..1969B,bib:PhReD1993..48..1629S}, also
related to what we are going to discuss right below. A discussion in connection with
wormholes also appeared \cite{bib:NuPhy1989B325...619P}.},
so that it is rather natural to assume an upper bound for the curvature
\cite{bib:JETPL1982..36...265M,bib:PhLeA1983..94...427M,bib:AnnPh1984.155...333M,%
bib:JETPL1984..40..1043M,bib:PhLeA1984.104...200M,bib:NuCim1985..86....97M},
which in these early approaches was naturally taken at the Planck scale.
Under this hypothesis, it can be shown that the black hole region of the Schwarzschild
metric can be joined along a spacelike junction to de Sitter space, the transition happening
in a time of the order of the Planck time \cite{bib:PhLeB1989.216...272M,bib:PhReD1990..41...383M}.
The procedure can be carried out not only between the eternal, static spacetimes, but also considering
the black hole as a result of gravitational collapse; moreover, it is possible for the de Sitter space
to decay into a Friedman universe, from which the suggestive idea that a new universe might
be created from gravitational collapse arises. Concentrating a little bit more on the technical
aspects of this early proposal, in view of our future discussion we would like to point out
that, because of the spacelike character of the junction, the resulting spacetime contains
Cauchy horizons and, thus, is not globally hyperbolic; we also remark, that the spacelike
hypersurface at which the Kasner-like contraction of the Schwarzschild collapse is turned
into the deflation of de Sitter spacetime is followed by another hypersurface, where
the transition to an inflationary space takes place. This last surface is topologically
a three sphere ${\mathbb{S}} ^{3}$, so that the new world formed inside the black hole
is actually a closed one (see references quoted above). This not a coincidence,
if we remember our discussion of the Bardeen solution on page \pageref{top:BarSol}
in subsection \ref{subsec:barsol}. We also emphasize, for future reference, that the global
structure of spacetime substantially corresponds to the light-grayed area of the maximal
extension of the Bardeen solution of figure \ref{fig:pendiamuGmuc}. It is also interesting to observe that
it can be proved \cite{bib:PhReD1990..41...395P} that this model is stable in the following sense:
inside the black hole region the $T ^{t} _{t}$ component of the stress energy tensor can
be interpreted as a tension along the axis of a three-cylinder of constant time $r = \mathrm{const.}$;
this is true, in particular, along the surface at which the junction is performed and it can be seen
that there are values of the parameters for which fluctuations of the Schwarzschild mass $M$
and/or of the de Sitter cosmological constant $\Lambda$ and/or of the other internal parameters,
as the surface pressures, do not induce a collapse of the three cylinder to zero radius, but
give rise, instead, to spatial oscillations of its radius with a longitudinal dependence.
In a modern perspective, it is also interesting to remember a later extension of this
early model \cite{bib:PhReD1996..53..3215F}, discussing the possibility of creation
of multiple de Sitter universes with null boundaries inside the black hole region,
a picture very similar to the one of an eternally inflating universe.

Other kinds of matching, which do not involve the presence of a surface layer, can also
be performed: in \cite{bib:ClQuG1996..13...L51S} the mass
function is given in two explicit examples, where the junctions are performed at the horizon
of Schwarzschild spacetime, which represents the exterior asymptotically flat region of the
solutions. Complete extensions of the manifold are also discussed and they cannot be
obtained by analytical continuation, which is not possible across the horizon \cite{bib:ClQuG1993..10..1865S}.
As discussed in \cite{bib:ClQuG1996..13...L51S}, it is also possible to perform the junction away
from the horizon without changing the general properties of the result: still the spacetime manifold,
which can be extended to completeness, is not analytical, so that its extension is not unique.
More complicated models with similar properties can also be obtained.

\subsection{\label{subsec:exasol}Exact solutions}

\begin{figure}
\begin{center}
\fbox{\includegraphics[width=3cm]{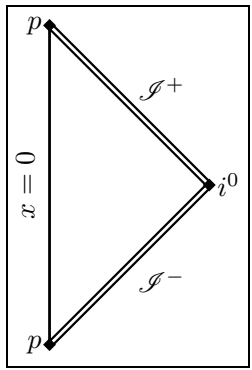}}
\caption{\label{fig:pendiamuLmuc}{\small{}generic spacetime structure for a G-lump.
In various regular solutions this structure arises from the same metric that gives rise to the extension
in figure {\protect\ref{fig:pendiamuGmuc}} but in a different range of the parameters (for instance,
with reference to the discussion in subsection \ref{subsec:maxext}, this structure it is obtained from the metric in
({\protect\ref{eq:adimetfor}}) and ({\protect\ref{eq:adimetfun}}) when $0 < \mu _{0} < \mu _{0} ^{\mathrm{cr}}$).
As before, conventions for names of boundaries follow the ones in \cite{bib:CaUPr1973...1...391E}.
We remember that spacetime is everywhere regular and that
it approaches de Sitter spacetime at small $r$ and Schwarzschild spacetime at large $r$.}}
\end{center}
\end{figure}

All the above models, with the exception of the Bardeen solution, are not exact solutions of
Einstein equations in the usual sense, since they do not satisfy everywhere the required
analyticity properties of the sources and/or the field equations. It is thus interesting to
consider if the Bardeen solution is just an isolated counterexample of can be obtained as a
particular case in a more general class of solutions. We will review, in the following
the instructive path followed in the literature to prove that the latter applies.
Seminal ideas in this direction (in the
framework of cosmology) actually appeared very early and much closer in time to the original
Bardeen solution than to related future developments: in particular already in the mid seventies
a \emph{regular} cosmological model was presented, where the idea that, with the increase of density matter
would enter a state dominated by a negative pressure \cite{bib:SovAL1975...1....93D}, was realized.
In this model a huge increase in the mass of the universe from the epoch of the beginning of expansion was obtained,
a result that suggestively anticipates some ideas that will be developed only years later, when regular
cosmological models will be formulated. Closer in spirit to the idea of curing future singularities
is, instead, the exact solution of Einstein equations presented in \cite{bib:GRGra1992..24...235D}.
In the notation introduced above in (\ref{eq:metfor}), this spherically symmetric static solution
of Einstein equations is characterized by
\begin{equation}
    f (r) = 1 - \frac{m (r)}{r} ,
    \quad \mathrm{where} \quad
    m (r) = r _{\mathrm{g}} \left( 1 - \exp \left( - \frac{r ^{3}}{r _{\ast} ^{3}} \right) \right)
    ;
\label{eq:dymmet}
\end{equation}
this model has, in fact, two free parameters which, in view of their physical meaning it is convenient
to chose as $r _{\mathrm{g}}$ and $r _{0} = \sqrt{r ^{3} _{\ast} / r _{\mathrm{g}}}$. It is not difficult
to check that if $r \gg r _{\ast}$ the metric resembles better and better the Schwarzschild solution, whereas
if $r \ll r _{\ast}$ a de Sitter like behavior is recovered \cite{bib:GRGra1992..24...235D}, since the metric
behaves as
\[
    f (r) \approx 1 - \frac{r _{\mathrm{g}}}{r _{\ast} ^{3}} r ^{2} = 1 - \frac{r ^{2}}{r _{0} ^{2}}
    \quad \mathrm{when} \quad
    r \approx 0
    .
\]
The stress-energy
tensor acting as source of this solution is such that $T ^{0} _{0} = T ^{1} _{1} \neq T ^{2} _{2} = T ^{3} _{3}$
(the last equality is a consequence of spherical symmetry). Following \cite{bib:SPJET1966..22...378G}
we see that in this model the source of the gravitational field is a realization of what can be called
a radial vacuum. It can be seen that the above model is regular everywhere; on the other hand,
for some values of the parameters an event horizon appears (which, as we will see later,
is then paired with a Cauchy horizon), so
that energy conditions have to be violated somewhere, and in fact, the strong energy condition is
certainly violated at small $r$, where the solution becomes de Sitter like. The weak energy condition
is instead satisfied everywhere. This fact shows that, when the parameters allow for the presence of
an event horizon in this regular spacetime, the spacetime cannot be globally hyperbolic. A detailed analysis of the analytic
extension in these situations \cite{bib:IJMPD1996...5...529D} confirms this educated guess (the global
structure of the solution is, again, as in figure \ref{fig:pendiamuGmuc}, showing an alternating sequence
of asymptotically flat Schwarzschild-like regions and regular de Sitter cores separated by black/white
hole regions). In view of the fact that there exist extremal configurations in which the Cauchy horizon
degenerates with the event horizon, it becomes particularly interesting to analyze the thermodynamic
properties of the solution\footnote{Interest for thermodynamic properties of regular spacetimes with horizons
appeared early \cite{bib:LNuCi1982..33...127G}, especially in connection with te idea that a remnant of some
kind is left at the end of the evaporation stage. See, for instance, \cite{bib:PhReL2006..96031103H} and
the references discussed in section \ref{sec:gausou} for recent examples.}; it is then possible to see that the temperature
drops to zero when the two horizons merge and the issue of the stability of the extremal configuration becomes
then of utmost importance to understand the final fate of the process \cite{bib:IJMPD1996...5...529D,bib:ClQuG2005..22..2331G}
(the stability of G-lumps is also being studied \cite{bib:PhLeB2007.645...358G}).
Although the solution described above is asymptotically flat, i.e. it realizes a transition from a non-zero
value of the cosmological constant at the center to a zero value at infinity, asymptotically de Sitter
generalizations can also be constructed \cite{bib:GRGra1998..30..1775S}: they are, of course, relevant
for cosmological applications\footnote{See, for instance, \cite{bib:ClQuG2003..20..3797D,bib:gr-qc2007..05..2368D}
and reference therein; we also, incidentally, note that in this context, the formation of baby universes inside
regular black holes has been also considered \cite{bib:PhLeB2001.506...351G}.}.
Also this family of solutions, as the one of the black hole type, can be
interpreted in a ``dynamical cosmological constant'' framework \cite{bib:PhLeB2000.472....33D}, but
we will not put the accent on this point in what follows. We think that, instead, it can be interesting to
remark the fact that the same solution, but in a different range of the relevant parameters, also describes
regular spacetimes without horizons, which are called G-lumps and represent a vacuum, self-gravitating,
regular, particle-like structure \cite{bib:ClQuG2002..19...725D,bib:PhLeB2007.645...358G},
similar in the spirit to boson stars \cite{bib:PhRev1968.172..1331K,bib:PhRev1969.187..1767B}, recently reviewed in
\cite{bib:NuPhy2000B564...185S,bib:ClQuG2004..21..1135W}). A representation of the global
structure of these solutions is given in figure \ref{fig:pendiamuLmuc}.
In this interpretative framework
it is found that the Arnovitt-Deser-Misner (ADM) mass is related to the cosmological term
\cite{bib:ClQuG2002..19...725D}, a result which remembers the remark we made at the very beginning
of this subsection and that is suggestive of interesting speculations.

\subsubsection{\label{subsubsec:matcon}Matter content}

The results summarized above can be understood in a consistent theoretical framework, which we will discuss
more precisely later on, in subsection \ref{subsec:genfra}.
Before that, in this subsection we would like to review how it is possible
to find physical motivations for matter content that generates solutions of the kind that we have
discussed in the previous subsections. The importance to provide a physical origin for the source
term that, \emph{via} Einstein equations, is then associated to regular black holes solutions,
has been emphasized since the earliest models that have appeared. We already discussed the very sound physical
motivation behind Sakharov \cite{bib:SPJET1966..22...241S} and Gliner \cite{bib:SPJET1966..22...378G} ideas. Moreover
the specific form of the solution (\ref{eq:dymmet}) has been motivated (see \cite{bib:IJMPD1996...5...529D}
and the references quoted there in the discussion on page 532) using the Schwinger formula \cite{bib:PhRev1951..82...664S},
according to which the probability $P _{\mathrm{Schwinger}}$ to find a particle as a result of vacuum polarization
in a given field $\Psi$ can be approximated by
\[
    P _{\mathrm{Schwinger}} \sim \exp ( - \Psi _{\mathrm{cr.}} / \Psi)
    .
\]
Using de Sitter spacetime with horizon radius $r _{0}$ (see above) to describe the properties of
the vacuum with polarization effects that manifest themselves as gravitational field tension,
we have
\[
    \Psi \sim \frac{r _{\mathrm{g}}}{r ^{3}}
    \quad \mathrm{and} \quad
    \Psi _{\mathrm{cr.}} = \frac{1}{r _{0} ^{2}}
    :
\]
in term of a parameter $r _{\mathrm{g}}$ with the dimension of a length,
$\Psi$ is the curvature term characteristic of the gravitational tension
and $\Psi _{\mathrm{cr.}}$ is the curvature critical value.
Thus
\[
    P _{\mathrm{Schwinger}} \sim \exp ( - r ^{3} / ( r _{0} ^{2} r _{\mathrm{g}}))
    ,
\]
which behaves as the $T ^{0} _{0}$ associated with the solution (\ref{eq:dymmet}) of
Einstein equations.

In view of the above analogy, it is interesting to consider more elaborate models,
in which it is possible to find some self-consistent solutions for the dynamics of gravity
and of the fields coupled to it, without assigning the form of the stress-energy tensor
\emph{a priori}\footnote{In a historical perspective, the interested reader can take a look at
the ideas present in \cite{bib:AnnPh1979.118....84S}, where, in a not too different conceptual
context, a similar problem was early investigated.}.
A framework in which the above program could be realized, is the one of nonlinear electrodynamics
(see\footnote{We emphasize the authors' names in this
particular case only, since in some references existing in the literature
there is a recurrent imprecision in the list of the authors of \cite{bib:PRSLA1934.143...410B}.}
Born \cite{bib:PRSLA1934.143...410B} and Born and Infled \cite{bib:PRSLA1934.144...425I})
coupled to gravity. With the above motivation
regular solutions in nonlinear electrodynamics theories coupled to gravity have been searched
\cite{bib:ClQuG1994..11..1469O,bib:PhReD1995..52..6178S,bib:PhLeB1998.432...287P}.
One example is given by the metric defined by the metric function \cite{bib:PhReL1998..80..5056G}
\[
    f (r) = 1 - \frac{2 m r ^{2}}{(r ^{2} + q ^{2}) ^{3/2}} + \frac{q ^{2} r ^{2}}{(r ^{2} + q ^{2}) ^{2}}
\]
for which the electric field, correspondingly, is obtained as
\[
    E (r)
    =
    q r ^{4}
    \left(
        \frac{r ^{2} - 5 q ^{2}}{(r ^{2} + q ^{2}) ^{4}}
        +
        \frac{15}{2} \frac{m}{(r ^{2} + q ^{2}) ^{7/2}}
    \right)
    ;
\]
this result is reminiscent of the Bardeen solution \cite{bib:PhLeB2000.493...149G},
of which it shares most of the properties.
In view of the discussion in the next section, we especially stress the asymptotic behaviors
of the solution, i.e.
\[
    f (r) \approx 1 - \left( \frac{2 m}{q} - 1 \right ) \frac{r ^{2}}{q ^{2}}
    , \quad
    E (r) \approx \frac{5 r ^{4}}{q ^{5}} \left( \frac{3}{2} \frac{m}{q} - 1 \right)
    \quad \mathrm{when} \quad
    r \approx 0
\]
and
\[
    f (r) \approx 1 - \frac{2 m}{r} + \frac{q ^{2}}{r ^{2}}
    , \quad
    E (r) \approx \frac{q}{r ^{2}}
    \quad \mathrm{when} \quad
    r \approx + \infty
    ;
\]
we see that, again, a de Sitter-like core occupies the small radius region, whereas at infinity
the solutions approaches the Rei\ss{}ner-Nordstr\o{}m solution, with the standard $r ^{-2}$
behavior of the electric field: $m$ and $q$ can then be interpreted as the mass and charge of the
solution and there are again values of the parameters for which the solution, which satisfies
the weak energy solution, is regular everywhere \cite{bib:PhReL1998..80..5056G} but has two horizons:
in this case the maximal extension again is given by the diagram in figure \ref{fig:pendiamuGmuc}.
Different realizations have also been found
\cite{bib:PhLeB1999.464....25G,bib:GRGra1999..31...629G,bib:PhReD2003..67124004B}.
A subsequent general analysis of these solutions was also performed \cite{bib:PhReD2001..63044005B}
and it was rigorously proved that regular electrically charged  structures are not compatible
with the Maxwell weak-field limit at the center; regular magnetic solutions are also analyzed
in \cite{bib:PhReD2001..63044005B}, under very general terms, as a viable alternative that respects
more closely some standard properties of the classical electromagnetic field.
A careful discussion of this and related issues can be found also in \cite{bib:ClQuG2004..21..4417D},
where another solution is introduced and its properties in relation with photon propagation
and the cosmological constant are analyzed. Another multi-parametric solution was recently discussed
in \cite{bib:GRGra2005..37...635G} and the spinning case has also bene studied \cite{bib:PhLeB2006.639...368D}.
All these solutions have the same/similar general properties of the ones
that we have discussed with some more detail.

\subsection{\label{subsec:genfra}The general framework}

\begin{figure}
\begin{center}
\fbox{\vbox{\vskip -3 mm\hbox{%
\fbox{\includegraphics[width=4.5cm]{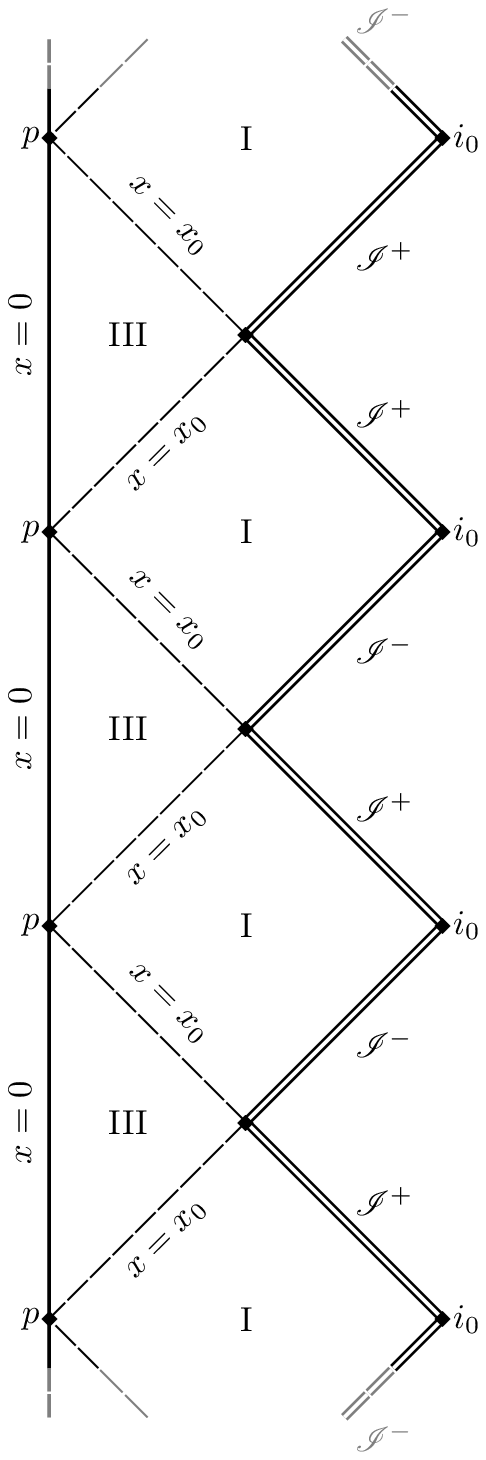}}\hskip 1 mm%
\parbox[b]{4.2cm}{\caption{\label{fig:pendiamuEmuc}{\small{}the maximal extension of the spacetime manifold for a black
hole with regular center when the two horizons (event and Cauchy) coincide (for the metric
defined in ({\protect\ref{eq:adimetfor}}) and ({\protect\ref{eq:adimetfun}}) this happens
when $\mu _{0} = \mu _{0} ^{\mathrm{cr}}$).  Again naming conventions of boundaries at infinity follow
the ones in \cite{bib:CaUPr1973...1...391E}, particularly we also indicated with $p$ the
exceptional points at infinity. Similar considerations can be made as in the other
cases shown before, in figures {\protect\ref{fig:pendiamuGmuc}} and {\protect\ref{fig:pendiamuEmuc}}.
In particular, the metric is again similar to Rei\ss{}ner-Nordstr\o{}m solution (this time to the
extremal one, of course), except again for the regular origin at $x = 0$ (i.e. $r = 0$).
The asymptotic behaviors
are also the same as the ones of the other cases, of course. Although we will not discuss
this point in the present work, extremal solutions are important when the thermodynamics of the
solution is considered, particularly in connection with the final stage of the evaporation
process.}}}}}}
\end{center}
\end{figure}

In the previous section we have seen (maybe in a not comprehensive, but still quite wide, perspective)
various solutions of Einstein equations that, despite the fact that they have event horizons,
are  not affected by the presence of singularities. We remember that we have restricted our attention
to the solutions which, in the classification given in \cite{bib:GRGra2007..39...973M}, are called
\emph{black holes with a regular center}. It turns actually out that these solutions can be (and have
been) discussed under a very general framework: it has, indeed, been proved that their nature is
essentially \emph{unique}, although their specific realizations may analytically differ.
We would like to discuss here in
more detail these interesting aspects, which relate the properties of the metric functions,
with those of the matter sources and the energy conditions that they satisfy. The constraints that
have to be satisfied by regular black hole solutions arise from singularities theorems, as well as from other
specific results (see, for instance, \cite{bib:ClQuG1993..10...327R,bib:PhLeA1996.218...147S}). In particular,
if we consider the case in which the matter is characterized by its energy density $\rho$, its
radial pressure $p _{\mathrm{r}}$ and its tangential pressure $p _{\mathrm{t}}$ (the presence of
only one non-radial pressure follows once we restrict ourself to the spherically symmetric
case, as we are doing here) it can be proved that, even in presence of junctions, the condition
\[
    \rho + p _{\mathrm{r}} + 2 p _{\mathrm{t}} \geq 0
\]
assures the absence of event horizons if the spacetime metric is $C ^{2}$ everywhere
\cite{bib:PhLeA1996.218...147S}: note that continuity of $\rho$ and $p _{\mathrm{t}}$
is not required. Thus, if we are interested in spacetimes which
are regular but have an event horizon, the condition written just above, must, in
general, be violated. It is very instructive to see that, in fact, black hole
solutions with a regular center can be characterized in a \emph{quasi}-complete way. This
has been done in \cite{bib:ClQuG2002..19...725D} (since the analysis in
\cite{bib:ClQuG2002..19...725D} is very clear and detailed we report here
only the final result, referring the reader to the original paper for the
interesting details of the proof, additional results and a careful discussion).
In particular it can be proved that if i) the dominant energy condition is
satisfied\footnote{The dominant energy conditions requires $T ^{00} \geq | T ^{ab} |$
($a , b = 1 , 2 , 3$); this condition is equivalent to \cite{bib:CaUPr1973...1...391E}
$\rho \geq 0${}$\:\:\wedge\:\:${}$- \rho \leq p _{k} \leq + \rho$ ($k = 1 , 2 , 3$), where
$\rho$ is the energy density and $p _{k}$ ($k = 1 , 2 , 3$) the principal pressures; in spherical symmetry
this is furthermore equivalent to $\rho \geq 0${}$\:\:\wedge\:\:${}$- \rho \leq p _{\mathrm{r}} \leq + \rho${}
$\:\:\wedge\:\:${}$- \rho \leq p _{\mathrm{t}} \leq + \rho$,
where $p _{\mathrm{r}}$ is the radial pressure and $p _{\mathrm{t}}$ the tangential pressure.}, ii) the metric is regular
at the center (and everywhere else), iii) the energy density is regular at the center,
iv) the solution is asymptotically
flat with finite ADM mass $M$ and v) the weak energy condition\footnote{The weak energy condition
requires $T _{\mu \nu} \xi ^{\mu} \xi ^{\nu} \geq 0$ for any timelike vector $\xi ^{\mu}$; this condition
is equivalent to \cite{bib:CaUPr1973...1...391E}
$\rho \geq 0${}$\:\:\wedge\:\:${}$\rho + p _{k} \geq 0$ ($k = 1 , 2 , 3$), where
$\rho$ is the energy density and $p _{k}$ ($k = 1 , 2 , 3$) the principal pressures; in spherical symmetry
this is equivalent to
$\rho \geq 0${}$\:\:\wedge\:\:${}$\rho + p _{\mathrm{r}} \geq 0${}$\:\:\wedge\:\:${}$\rho + p _{\mathrm{t}} \geq 0$,
where $p _{\mathrm{r}}$ is the radial pressure and $p _{\mathrm{t}}$ the tangential pressure.
Note that the weak energy condition is implied by the dominant energy condition, although not equivalent to it.}
is satisfied in some region around the origin $r = 0$, then $T ^{t} _{t} = T ^{r} _{r}$ and
\begin{eqnarray}
    f (r) = 1 - \frac{2 m (r)}{r}
    , &
    m (r) = 4 \pi {\displaystyle{}\int  _{0} ^{r} x ^{2} \rho (x) d x} \:\: \mathrm{with}
    &
    \lim _{r \to + \infty} m (R) = M < \infty
    ;
    \nonumber \\
    \mathrm{moreover} \: \:
    {\displaystyle\frac{d \rho}{d r} \leq 0},
    & p _{\mathrm{r}} = - \rho \quad \mathrm{and} \quad p _{\mathrm{t}} = - \rho - {\displaystyle\frac{r}{2}\frac{d \rho}{d r}} &
    \mathrm{which,\ in\ turn,\ imply}
    \nonumber
\end{eqnarray}
\label{top:theore}that the solution behaves asymptotically as de Sitter space when $r \to 0$. A straightforward corollary
of the above results \cite{bib:ClQuG2002..19...725D} is that, $f (r)$ has a maximum at $r = 0$ and
can have at most one more extremum that must be a minimum: then the metric can at most have two zeroes,
so at most two horizons. This happens when the minimum is negative; when the minimum
is exactly zero the two horizons degenerate and when the minimum is positive there are no horizons.
The possible causal structures of this class of solutions are reported in figures
\ref{fig:pendiamuGmuc}, \ref{fig:pendiamuLmuc} and \ref{fig:pendiamuEmuc}.
We remark the following points.
\begin{enumerate}
\item The dominant energy conditions is required
only to obtain the regularity of pressures from the assumption of the regularity of the energy density. If
regularity of the pressures in instead postulated, it is enough to assume the weak energy condition. In this
sense, the weak energy condition singles out the class of regular spherically symmetric metrics, which are
asymptotically de Sitter at the center and asymptotically Schwarzschild at infinity; the results about the
behavior of the metric function are also a consequence of the validity of the weak energy condition
\cite{bib:ClQuG2002..19...725D}.
\item In view of the general results discussed above the matter content associated to this class of solutions
describes an anisotropic fluid since $p _{\mathrm{r}} = p _{\mathrm{t}}$ can be satisfied if and only if
$\rho = \mathrm{const}.$; this implies that the only way to have isotropic solutions with a finite ADM mass is to
consider vacuum solutions. Moreover, as seen before \cite{bib:PhLeA1996.218...147S}, these solutions will
generically have
\[
    \rho + p _{\mathrm{r}} + 2 p _{\mathrm{t}} < 0
\]
and will violate the strong energy condition\footnote{The strong energy condition requires
$(T _{\mu \nu} - T g _{\mu \nu} / 2 ) \zeta ^{\mu} \zeta ^{\nu} \geq 0$ for all future-directed
timelike vectors $\zeta ^{\alpha}$ (T is the trace of $T _{\mu \nu}$). This is equivalent to the requirements
$\rho + p _{1} + p _{2} + p _{3} \geq 0${}$\:\:\wedge\:\:${}$\rho + p _{k} \geq 0$ ($k = 1 , 2 , 3$), where
$\rho$ is the energy density and $p _{k}$ ($k = 1 , 2 , 3$) the principal pressures; in presence
of spherical symmetry this is equivalent to
$\rho + p _{\mathrm{r}} + 2 p _{\mathrm{t}}\geq 0${}$\:\:\wedge\:\:${}$\rho + p _{\mathrm{r}} \geq 0${}$\:\:\wedge\:\:${}$\rho + p _{\mathrm{t}} \geq 0$,
where $p _{\mathrm{r}}$ is the radial pressure and $p _{\mathrm{t}}$ the tangential pressure.} somewhere.
\item When the solution has two horizons, the larger one is an event horizon, whereas
the smaller one is a Cauchy horizon, so that the solution is not globally hyperbolic (see, again,
figure \ref{fig:pendiamuGmuc}).
\item The hypothesis about asymptotic flatness can be relaxed, to allow the presence
of a cosmological constant; this aspect has also been extensively discussed
\cite{bib:GRGra1998..30..1775S,bib:IJMPD2003..12..1015D,bib:ClQuG2003..20..3797D}.
\end{enumerate}
Other general results about solutions with a de Sitter core have appeared in the literature
(see for instance \cite{bib:RepMP1999..44...407M}, where an equation of state of the
\emph{elastic} type \cite{bib:GRGra1992..24...139K} is considered, as well as the analysis
in \cite{bib:ClQuG2002..19..4399G}); extensions of the above results with
gravity coupled to a nonlinear scalar field are also very interesting (see, e.g.,
\cite{bib:PhReD2001..64064013B,bib:ClQuG2001..18..1715L} and references therein).
This general analysis is of course important and preliminary for physical application
of these ideas, which also have appeared (see, for instance, \cite{bib:IJMPD2003..12..1787D}).

\subsection{\label{subsec:genanasyn}Synopsis}

We have, thus, reviewed that black hole with a regular center can be characterized,
in a very precise way, in the context of general relativity; solutions which are
spherically symmetric, everywhere regular, satisfy the weak energy condition and
are asymptotically Schwarzschild have known global structures (as in figures
\ref{fig:pendiamuGmuc}, \ref{fig:pendiamuLmuc} and \ref{fig:pendiamuEmuc});
if they have an event horizon, they must have a Cauchy horizon (and they
admit the limit in which the two coincide): in these cases they are not
globally hyperbolic. Non-essential details of particular solutions are specified
by a unique function, the energy density, which has to be a non-increasing function
tending to zero fast enough at large distance. In all generality the matter source
will be an anisotropic fluid, which realizes a radial vacuum structure and
violates somewhere the strong energy condition. We feel to remark the nice
\emph{travel of knowledge} that, starting from the earliest idea about the
state of matter at very high densities, could find its way in the thin
space left open by singularity theorems, to obtain some particular
realization of singularity avoidance, first, and their general characterization,
later.

\section{\label{sec:gausou}Gaussian sources}

Recently another neutral realization of a black hole with a regular center has been derived
\cite{bib:PhLeB2006.632...547S} and then generalized to the charged case \cite{bib:PhLeB2007.645...261S}.
With the notation that we set up above, the neutral solution can be described by the following
choice of the metric function:
\begin{equation}
    f (r) = 1 - \frac{2 m (r)}{r}
    , \quad
    m (r) = \frac{2 m _{0}}{\pi ^{1/2}} \gamma \left( \frac{3}{2} , \frac{r ^{2}}{4 \Theta} \right)
    .
\label{eq:metfun}
\end{equation}
We will restrict ourself to the solution interpreted as a solution of classical Einstein equations
without entering into a discussion of the, also interesting, motivations for it,
which will be reviewed elsewhere by one of the authors of \cite{bib:PhLeB2006.632...547S}.
For our subsequent analysis it is enough to remember that the metric in equation (\ref{eq:metfun}) is
obtained assuming a Gaussian shaped, radial, mass energy density with total mass $m _{0}$
and width proportional to $\sqrt{\Theta}$: the property $T ^{t} _{t} = T ^{r} _{r}$
is also assumed, so that from the above results we know that the metric is a realization
of a black hole with regular center.

To streamline the following analysis, we will preliminarily make the change of variables
$r = \sqrt{2 \Theta} \, x$, $t = y / \sqrt{2 \Theta}$: in the coordinate system
$(y , x , \vartheta , \varphi)$ the metric, then, takes the form
\begin{equation}
    g _{\mu \nu}
    =
    {\mathrm{diag}}
    \left( - f(x) / 2 \Theta , 2 \Theta f (x) ^{-1} , 2 \Theta x ^{2} , 2 \Theta x ^{2} \sin ^{2} \vartheta \right)
\label{eq:adimetfor}
\end{equation}
where
\begin{equation}
    f (x) = 1 - \frac{2 \mu (x)}{x}
    , \quad
    \mu (x) = \frac{2 \mu _{0}}{\pi ^{1/2}} \gamma \left( \frac{3}{2} , \frac{x ^{2}}{2} \right) = \frac{m (r(x))}{\sqrt{2 \Theta}}
    , \quad \mu _{0} = \frac{m _{0}}{\sqrt{2 \Theta}}
    .
\label{eq:adimetfun}
\end{equation}
Note that, with the metric in this form, $x$ does not represent the circumferential radius, which is instead given by $\sqrt{2 \Theta} \, x$.
Moreover, both $x$ and $\mu _{0}$ are dimensionless parameters and $\mu (x)$ is a dimensionless function. The total
mass energy seen by an observer at asymptotic infinity is $m _{0} = \sqrt{2 \Theta} \, \mu _{0}$, whereas $m(r) = \sqrt{2 \Theta} \, \mu (x)$
is the total mass energy inside the radius $r = \sqrt{2 \Theta} \, x$.

With the intent of making some acquaintance with the parameters of the model, we will reproduce here some general results for the metric
in terms of the dimensionless set up described above. These properties can be analyzed directly from the study of
the dimensionless metric function $f (x) = 1 - 2 \mu (x) / x$
and can be conveniently expressed in terms of the two parameters $\mu _{0}$, already defined in (\ref{eq:adimetfun}),
and $x _{0}$, which is the only positive root of the equation\footnote{The existence of one and only one
positive root comes from the general results discussed in subsection \ref{subsec:genfra}, given that the
equation below is equivalent to $f ' (x) =  0$; anyway, this property can also be easily verified directly.}
\[
    2 e ^{- x ^{2} / 2} \frac{x}{\sqrt{2}} = \frac{\gamma ( 3 / 2 , x ^{2} / 2)}{x ^{2}/2}
    .
\]
In particular,
\[
    f ' ( x ) = 0 \quad \Leftrightarrow \quad x = x _{0} \: \vee \: x = 0
\]
and
\[
    f '' ( 0 ) < 0 , \quad f '' (x _{0}) > 0
\]
so that $x = 0$ is a local maximum and $x = x _{0}$ is a local minimum.
At these points the values of the metric function are
\[
    f ( 0 ) = 1 , \quad f ( x _{0} ) = 1 - \frac{\mu _{0}}{\mu _{0} ^{\mathrm{cr}}}
    , \quad \mathrm{where} \quad
    \mu _{0} ^{\mathrm{cr}} = \frac{\pi ^{1/2}}{4 \sqrt{2}} \frac{e ^{x _{0} ^{2} / 2}}{x _{0} ^{2} / 2}
    ;
\]
from the additional result $\lim _{x \to + \infty} \mu (x) = 1 ^{-}$,
which follows remembering the definition of $\mu (x)$ in (\ref{eq:adimetfun}) and the identity
\begin{equation}
    \lim _{y \to + \infty} \frac{2}{\pi ^{1/2}} \gamma \left( \frac{3}{2} , y \right )
    =
    \frac{2}{\pi ^{1/2}} \Gamma \left( \frac{3}{2} \right ) = 1
    ,
\label{eq:gampro}
\end{equation}
it turns out that three cases can happen:
\begin{description}
    \item[$\mu _{0} > \mu _{0} ^{\mathrm{cr}})$]
    the metric function $f (x)$ vanishes at two points $x _{1} < x _{2}$;
    \item[$\mu _{0} = \mu _{0} ^{\mathrm{cr}})$] the metric function has a double zero, since
    $f (x _{0}) = 0$ exactly at the minimum point $x _{0}$;
    \item[$0 \leq \mu _{0} < \mu _{0} ^{\mathrm{cr}})$] the metric function is always positive,
    so it has no zeroes.
\end{description}
In the case in which the metric function has two zeroes, clearly
we have $0 \leq x _{1} \leq x _{0}$. To locate (slightly) more precisely the
position of $x _{2}$, we remember the definition of $\mu (x)$ in (\ref{eq:adimetfun}), the property
(\ref{eq:gampro}) and that the lower incomplete gamma function is a monotonically
non-decreasing function of its second argument (since $d _{y} \gamma ( 3/2 , y ) = \sqrt{y} e ^{-y} \geq 0$).
Thus $\mu (x) \leq \mu _{0}$ and the difference between $\mu (x)$  and $\mu _{0}$ decreases to zero
as $x \to \infty$. Thus
\[
    \lim _{\mu _{0} \to + \infty} \frac{x _{2}}{2 \mu _{0}} = 1
\]
and always $x _{2} \leq 2 \mu _{0}$; in other words $2 \mu _{0}$ is an upper bound for the
position of $x _{2}$. Then, in general
\[
    0 \leq x _{1} \leq x _{0} \leq x _{2} \leq 2 \mu _{0}
    .
\]
A numerical evaluation gives $x _{0} \approx 2.1372$ and $\mu _{0} ^{\mathrm{cr}} \approx 1.3464$.
For future reference, we also remember that the characteristic scale $r \sim \sqrt{2 \Theta}$,
corresponds to $x \sim 1$. The above results are a dimensionless paraphrase of the one present
in \cite{bib:PhLeB2006.632...547S}, which will turn out to be convenient later on; they realize in
the present particular case the general framework described in subsection \ref{subsec:genfra}.

\subsection{\label{subsec:maxext}Maximal extension}

Using the properties detailed above, we will now reproduce in this particular case the general results about
the maximal extension of the solution. Since the timelike two surface that we obtain ignoring the spherically
symmetric component, ${\mathbb{S}} ^{2}$, of the spacetime manifold satisfies all the requirements
of the diagrammatic extension procedure described in ref. \cite{bib:JMaPh1970..11..2280W}
we are going to follow this immediate procedure. Three cases have then to be analyzed, corresponding to
the presence of two, one or no zeroes of the metric function $f (x)$.
\begin{figure}
\begin{center}
\fbox{\vbox{\hbox{\includegraphics[width=11.6cm]{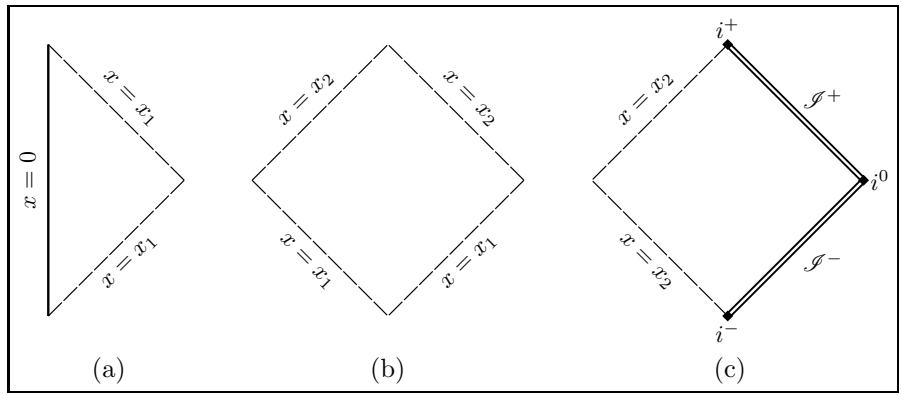}}\hbox{\hskip 1.0 cm{}(a)\hskip 3.25 cm{}(b)\hskip 4.1 cm{}(c)}}}
\caption{\label{fig:maxextmuGmuc}{\small{}blocks to be joined to obtain the maximal extension of the metric in
(\ref{eq:adimetfor}), (\ref{eq:adimetfun}), when we have $\mu _{0} > \mu _{0} ^{\mathrm{cr}}$.
Three blocks can be constructed, corresponding (a) to a timelike region between $x = 0$ and $x = x _{1}$,
(b) to a spacelike region between $x = x _{1}$ and $x = x _{2}$ and (c) to another timelike region between $x = x _{2}$
and infinity. See also the main text.}}
\end{center}
\end{figure}
Conventions for the
labels of boundaries, which in this case are infinities, follow ref. \cite{bib:CaUPr1973...1...391E},
but note that a double line represents asymptotically flat null infinity.

\paragraph{Metric function with two zeroes at $x _{1}$ and $x _{2}$.}
\label{pag:maxext001}In this case the block
extension to obtain the full spacetime manifold goes as follows. The two zeros divide the
$[0 , + \infty)$ interval of variation of the $x$ coordinate in three regions. The block
corresponding to the first of them is the one labelled (a) in figure \ref{fig:maxextmuGmuc}
and covers the region between $x = 0$ and $x = x _{1}$; $x = 0$ is timelike and
constant-$x$ lines join the bottom vertex with the top one. A second block (labelled (b) in
figure \ref{fig:maxextmuGmuc}) covers the region
between $x _{1}$ and $x _{2}$. Now constant $x$-curves are in the opposite direction as
compared with the previous block, since the value of the metric function in this region
is non-positive, which corresponds to exchanging the timelike and spacelike directions:
they thus join the left and right vertices of the corresponding block.
Finally the third block, labelled (c) in figure \ref{fig:maxextmuGmuc},
covers the region between $x _{2}$ and infinity, has again
$f (x) \geq 0$ and the constant-$x$ lines are directed as in the first diagram, so they join
$i ^{-}$ and $i ^{+}$. The metric is asymptotically flat and we represent this fact in the
diagram with the doubled lines at null future and past infinity.
The maximally extended spacetime is obtained by sewing together the blocks described above \cite{bib:JMaPh1970..11..2280W}.
This gives the result in figure \ref{fig:pendiamuGmuc} on page \pageref{fig:pendiamuGmuc}.
The global extension is Rei\ss{}ner-Nordstr\o{}m like, but note that $x = 0$
is not a singularity.
A point in the diagram, at position $x$, represents a
$2$-sphere ${\mathbb{S}} _{2}$ of radius $\sqrt{2 \Theta} \, x$. Dashed lines represent values of $x$
at which $f (x) = 0$ and they are null lines. The maximal extension has been, in fact,
conformally mapped so that light rays travel along lines forming $\pm 45$ degrees with the
horizontal direction.

\begin{figure}
\begin{center}
\fbox{\vbox{\hbox{\includegraphics[width=7cm]{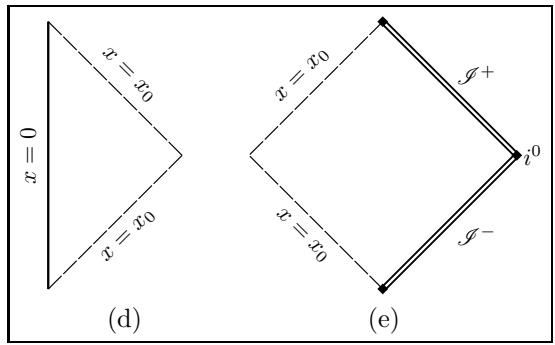}}\hbox{\hskip 1.2 cm{}(d)\hskip 3 cm(e)}}}
\caption{\label{fig:maxextmuEmuc}{\small{}blocks to be joined to obtain the maximal extension of the metric
({\protect\ref{eq:adimetfor}}), ({\protect\ref{eq:adimetfun}})
when we have $\mu _{0} = \mu _{0} ^{\mathrm{cr}}$. Two blocks can be constructed,
corresponding to a timelike region between $x = 0$ and $x = x _{1}$ (d) and another timelike
region between $x = x _{1}$ and infinity (e). The maximally extended manifold is shown
in figure {\protect\ref{fig:pendiamuEmuc}} on page {\protect\pageref{fig:pendiamuEmuc}}.}}
\end{center}
\end{figure}
\paragraph{Metric function with one zero at $x _{0}$.} Although the result will not be a surprise,
for completeness we analyze also this case, which corresponds to $\mu _{0} = \mu _{0} ^{\mathrm{cr}}$;
again we find that the maximally extended diagram is of the Rei\ss{}ner-Nordstr\o{}m type (extremal,
of course) and, again, with a regular origin at $x = 0$. The extension procedure can be graphically performed as before.
The relevant blocks are now two, shown in figure \ref{fig:maxextmuEmuc}.
The first of them (labelled (d) in figure \ref{fig:maxextmuEmuc}) covers the patch between
$x = 0$ and $x = x _{0}$: at $x = x _{0}$ the metric function
and its first derivative vanish (dashed line). The second (labelled (e) in figure \ref{fig:maxextmuEmuc})
covers the patch between $x = x _{0}$ and infinity, which
is asymptotically flat (the double lines are, as before past and future null infinity). Combining
this diagrams in all possible ways, so that the metric in the resulting manifold is regular
across the sews, gives, as a final result, the maximal extended diagram in Fig. \ref{fig:pendiamuEmuc}.
Again, this is a well know infinite sequence of the patches described above, except for
the fact that $x = 0$ is regular (concerning the structure of the diagram and the representation of light cones,
the same technical details apply, as in the previous case of Fig. \ref{fig:pendiamuGmuc}).

\paragraph{Metric function with no zeros.} We deal quickly with the last case, which
corresponds to\footnote{We are excluding the case $\mu _{0} = 0$, which corresponds
to Minkowski spacetime.} $0 < \mu _{0} < \mu _{0} ^{\mathrm{cr}}$. Now the metric never
vanishes and only one block is obtained, which also represents the maximal extension
of the metric in this range of variation of $\mu _{0}$ (the result is as in figure
\ref{fig:pendiamuLmuc}). The origin is again regular,
so that we obtain a Minkowski--like structure, except that spacetime is not empty in
this case. If we consider this result in conjunction with the two previous
cases, we can complete the analogy with the Rei\ss{}ner-Nordstr\o{}m solution that we
repeatedly emphasized: note that also in this case we have a regular origin instead that
the Rei\ss{}ner-Nordstr\o{}m timelike, naked singularity.

We have thus reproduced in this particular realization the general result discussed in
subsection \ref{subsec:genfra}, i.e. the fact that the maximal extension, geometrically,
``looks like'' the Rei\ss{}ner-Nordst\o{}m solution but with a regular origin. This, non minor, difference
makes the spacetime manifold everywhere well behaved. After this \emph{detour} to explicitly discuss the global
geometry of the spacetime, we will turn to its energy matter content in the following section.

\subsection{\label{subsec:enecon}Energy conditions}

\begin{figure}
\begin{center}
\fbox{\includegraphics{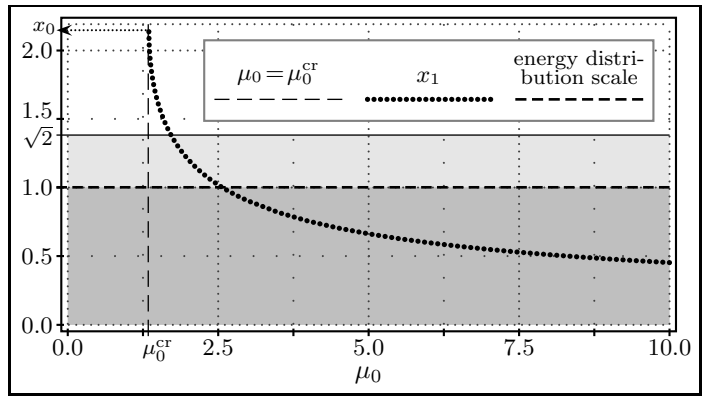}}
\caption{\label{fig:StrEneCon}{\small{}this diagram compares, for different values of $\mu _{0}$,
the position of the Cauchy horizon, the size of the region violating the strong energy
condition (\emph{union} of light and dark gray area) and the region corresponding to the
characteristic size of the Gaussian energy distribution which is the source of
the solution (horizontal dashed line). We see that the size of the energy violating region and the characteristic
size of the energy distribution are of comparable magnitude and that, both, extend beyond
the Cauchy horizon for large enough values of $\mu _{0}$.}}
\end{center}
\end{figure}

Up to now we have discussed geometrically the spacetime associated with the metric
(\ref{eq:adimetfor}), (\ref{eq:adimetfun}) as a representative of the more general class of black
hole spacetime with a regular center, without any reference to the matter content
that generates the solution. We will briefly review here the properties of the sources;
since the metric (\ref{eq:adimetfor}), (\ref{eq:adimetfun}) has
been derived elsewhere, we will take here the fast way, and just calculate
the Einstein tensor associated to the solution. This is given by
\begin{equation}
    G ^{\mu} _{\nu}
    =
    \frac{1}{\sqrt{2 \Theta}}
    \mathrm{diag}
    \left( - \frac{2 \mu ' (x)}{x ^{2}} , - \frac{2 \mu ' (x)}{x ^{2}} , - \frac{\mu '' (x)}{x} , - \frac{\mu '' (x)}{x} \right);
\end{equation}
so that the corresponding stress energy tensor is
\begin{equation}
    T ^{\mu} _{\nu}
    =
    \frac{1}{8 \pi} G ^{\mu} _{\nu}
    =
    \frac{1}{8 \pi \sqrt{2 \Theta}}
    \mathrm{diag}
    ( - \epsilon (x) , p _{x} (x) , p _{\mathrm{t}} (x) , p _{\mathrm{t}} (x) )
    ,
\end{equation}
where $\epsilon (x)$ is the energy density, $p _{x} (x)$ the radial pressure and
$p _{\mathrm{t}} (x)$ the tangential pressure. Explicitly,
\begin{eqnarray}
    \epsilon (x)
    & = &
    \frac{2 \mu ' (x)}{x ^{2}}
    =
    \frac{\mu _{0}}{\sqrt{2 \Theta} (2 \pi) ^{3/2}} e ^{-x ^{2} / 2}
    \nonumber \\
    p _{x} (x)
    & = &
    - \frac{2 \mu ' (x)}{x ^{2}}
    =
    - \epsilon (x)
    \\
    p _{t} (x)
    & = &
    - \frac{\mu '' (x)}{x}
    =
    -
    \frac{\mu _{0}}{\sqrt{2 \Theta} (2 \pi) ^{3/2}}
    e ^{- x ^{2} / 2} \left( 1 - \frac{x ^{2}}{2} \right)
    .
    \nonumber
\end{eqnarray}
\begin{figure}
\begin{center}
\fbox{\includegraphics{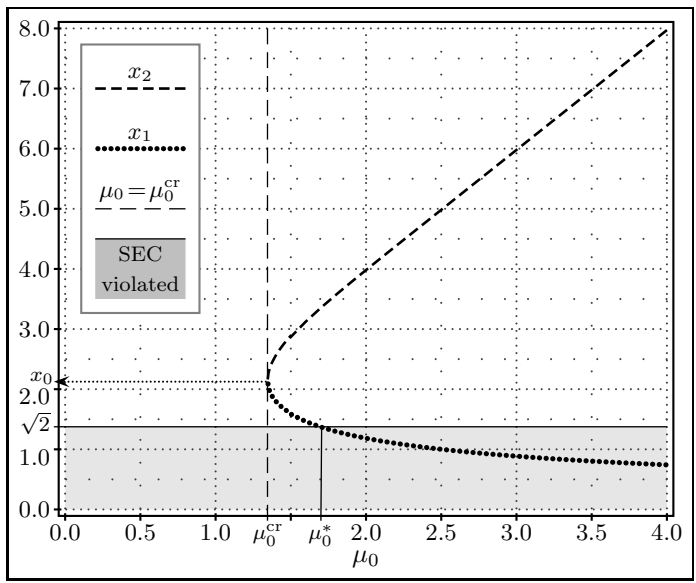}}
\caption{\label{fig:eneviopos}{\small{}plot i) of the position of the two zeroes,
$x _{1}$ and $x _{2}$, of the dimensionless metric function and ii) of the position
of the dimensionless radius below which the strong energy conditions is violated
as functions of $\mu _{0}$ (the second quantity is constant in the chosen units
and corresponds to the horizontal line with ordinate $\sqrt{2}$). The grayed area
shows where the strong energy condition is violated. This happens a) for the solutions
of the G-lump kind, i.e. the ones with no horizons which have $0 < \mu _{0} < \mu _{0} ^{\mathrm{cr}}$,
b) for the extremal solutions inside the degenerate horizon and for the two horizon solutions
c1) inside the Cauchy horizon or c3) between the two horizons or c2) until the inner
Cauchy horizon, depending on the value of $\mu _{0}$ and if it is smaller, bigger
or equal than $\mu ^{\ast}$ (the itemization of the various
possibilities follows the one in the main text).}}
\end{center}
\end{figure}
From the above results it is easy to see that \emph{the weak energy condition
is always satisfied}, since
\[
    \epsilon (x) \geq 0 , \quad
    \epsilon (x) + p _{x} (x) = 0 \geq 0 , \quad
    \mathrm{and} \quad
    \epsilon (x) + p _{t} (x) \sim x ^{2} e ^{- x ^{2} / 2} \geq 0
    .
\]
This again matches with the general properties of black holes with a regular
center discussed in subsection \ref{subsec:genfra}. We are now interested
to consider the strong energy condition, which we know has to be violated,
to determine the qualitative properties of this violation. We would like, in
particular, to identify where the violation can occur. With pure algebra we
obtain
\[
    \epsilon ( x ) + p _{x} (x) + 2 p _{t} (x)
    =
    2 p (t) (x)
    \sim
    e ^{- x ^{2} / 2} (x ^{2} - 2)
\]
so that the matter which is the source in the present solution violates
the strong energy condition for $0 \leq x < \sqrt{2}$ (the choice of scale
for the various coordinates allows for a very immediate expression of this
result, in which no free parameters appear). A handy way
to understand where, in the full spacetime manifold, the violation
occurs, is to plot, at once, the position of the horizons and of the
border of the strong energy condition violating region. This is the content
of figure \ref{fig:eneviopos}.
The zeroes of the metric function, corresponding to the points $x _{1}$
and $x _{2}$ if $\mu _{0} > \mu _{0} ^{\mathrm{cr}}$ (and that we called
$x _{0}$ in the limiting case in which they coincide, i.e. when
$\mu _{0} = \mu _{0} ^{\mathrm{cr}}$)
are plotted as functions of $\mu _{0}$. We recognize some of the features
already discussed above, as the presence of the critical value
$\mu _{0} ^{\mathrm{cr}}$ below which no zeroes of the metric function
appear; moreover the grayed area represents the values of the dimensionless
radial coordinate $x$ at which the strong energy condition is violated.
\begin{figure}
\begin{center}
\fbox{\includegraphics[width=2.8cm]{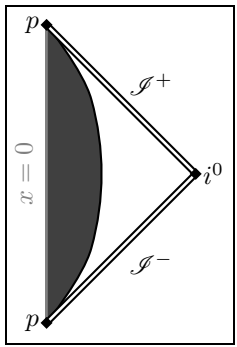}}
\caption{\label{fig:StrEneConVio004}{\small{}qualitative representation in the maximally
extended diagram of a solution with no horizons (i.e with $0 < \mu _{0} < \mu _{0} ^{\mathrm{cr}}$)
of the region where the strong energy condition is violated (dark gray area). In this solutions
a possible naked singularity (which, for instance, can be found in the geometrically analogous
Rei\ss{}ner-Nordstr\o{}m case) is replaced by a strong energy condition violating region. This is case
a) in the main text.}}
\end{center}
\end{figure}
We then see that:
\begin{enumerate}
\item[a)] the strong energy condition is violated ``around the regular origin''
for solutions which do not posses horizons
(i.e. the G-lump solutions of subsection \ref{subsec:exasol});
this can be seen from the part of the diagram on the left of the vertical
dashed line, i.e. for $0 < \mu _{0} < \mu _{0} ^{\mathrm{cr}}$ and is represented
by the dark gray area in the maximally extended diagram of figure \ref{fig:StrEneConVio004}.
\item[b)] in the extremal case, the strong energy condition is violated for values
of $x$ which are completely inside the region $x < x _{0}$, i.e. inside the
degenerate horizon; this is transparent from figure \ref{fig:eneviopos} and represented,
again with a dark gray shading, in the maximally extended diagram of figure
\ref{fig:StrEneConVio003}.
\begin{figure}
\begin{center}
\fbox{\includegraphics[width=4.5cm]{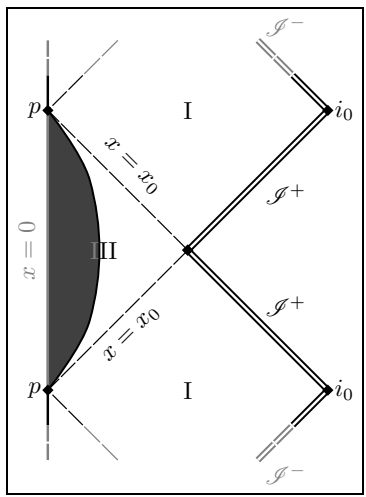}}
\caption{\label{fig:StrEneConVio003}{\small{}representation of the region in which the strong
energy condition is violated in the case in which $\mu _{0} = \mu _{0} ^{\mathrm{cr}}$, i.e.
when we have an extremal black hole with a regular center.
As seen from figure {\protect\ref{fig:eneviopos}} the
violation occurs inside the degenerate horizon $x _{0}$. The dark gray shaded region is
a qualitative representation of the region of spacetime affected by this violation. This is
case b) in the main text.}}
\end{center}
\end{figure}
\item[c)] in the non-extremal case, the strong energy conditions can be violated
in three qualitatively different situations, which can also be identified starting
from figure \ref{fig:eneviopos}:
\begin{enumerate}
\item[c1)] in a first case, for values of $\mu _{0}$ bigger than
$\mu _{0} ^{\mathrm{cr}}$ but smaller than $\mu _{0} ^{\ast}$
(we name in this way the value of $\mu _{0}$ at which the inner Cauchy horizon
is located at $x = \sqrt{2}$, i.e. at which the black dotted line crosses
the constant $\sqrt{2}$ line in figure \ref{fig:eneviopos})
the strong energy condition violating region is completely
confined inside the region $x < x _{1}$, i.e. inside the inner Cauchy horizon;
this is shown in figure \ref{fig:StrEneConVio002}, panel c1), with a dark gray region, again;
\begin{figure}
\begin{center}
\vbox{\hbox to 12.5cm {\fbox{\includegraphics[width=6cm]{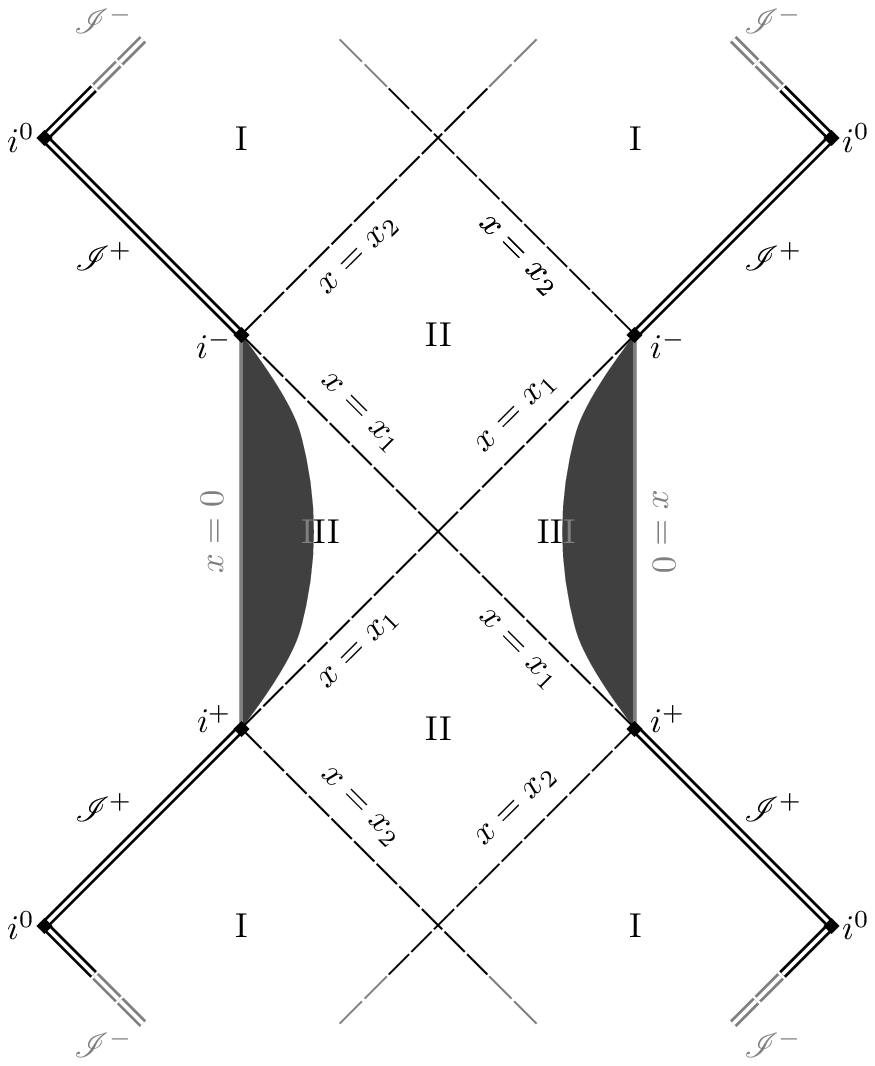}}%
\hfil%
\fbox{\includegraphics[width=6cm]{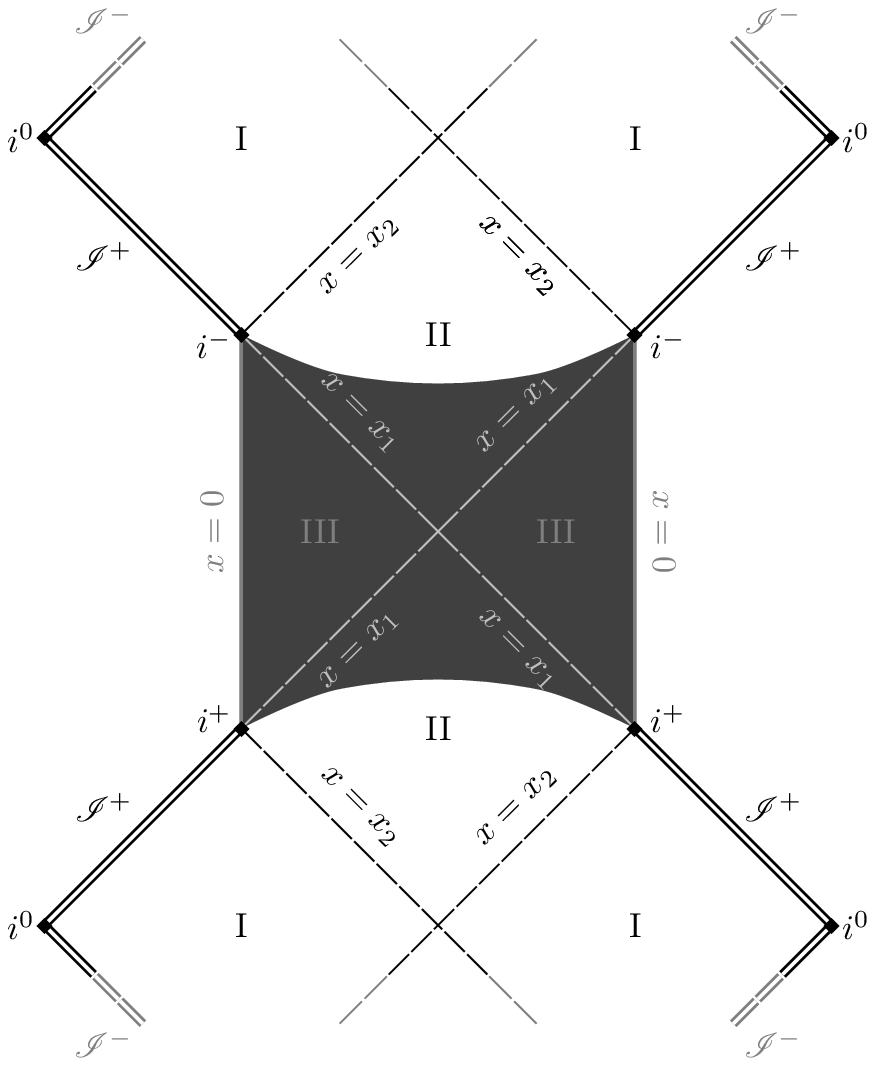}}}
\vskip 1 mm%
\hbox to 12.5cm {\hfil{}c1){}\hfil\hfil{}c3)\hfil}}
\caption{\label{fig:StrEneConVio002}{\small{}representation of the strong energy condition
violating region in the case in which $\mu _{0} > \mu _{0} ^{\mathrm{cr}}$, so that
the maximally extended spacetime has the structure shown in figure \ref{fig:pendiamuGmuc}
on page \pageref{fig:pendiamuGmuc}. As discussed in the main text, three situations
can happen: only cases c1) and c3 are shown here (items have the same name as in the
discussion of the main text). Case c1) corresponds to the case
$\mu _{0} ^{\mathrm{cr}} < \mu _{0} < \mu _{0} ^{\ast}$ and the strong energy
condition is violated in a region completely contained inside the cauchy horizon.
In the case c3), i.e. for $\mu _{0} > \mu _{0} ^{\ast}$, the strong energy condition violating
region includes, instead, all the spacetime inside the Cauchy horizon and part of the region
between the Cauchy horizon and the event horizon. The in between c2) case (not shown)
takes place when the strong energy condition is violated exactly inside the Cauchy horizon
region. As in the previous diagrams, follow the convention to show in dark
gray color the part of spacetime where the strong energy condition is violated:
the representation is qualitatively faithful.}}
\end{center}
\end{figure}
\item[c2)] in an intermediate case, with $\mu _{0} ^{\mathrm{cr}} = \mu _{0} ^{\ast}$,
the strong energy condition violating region is exactly the region
$x \leq x _{1}$;
\item[c3)] in a final third case, when $\mu _{0} > \mu _{0} ^{\ast} > \mu _{0} ^{\mathrm{cr}}$,
the strong energy condition violating region includes all the spacetime inside the Cauchy horizon
and part of the region between the Cauchy and event horizon, up to a value of the radius that
depends from the specific value of $\mu _{0}$. With the same conventions of the previous
pictures this is shown in figure \ref{fig:StrEneConVio002}, panel c3).
\end{enumerate}
\end{enumerate}
Notice that the above results are independent of the scale $\sqrt{2 \Theta}$, which,
nevertheless, has to be considered when passing from the dimensionless radial coordinate
$x$ to physical ones, as for instance the circumferential radius $r$. The analysis
identifies the region of spacetime where the violation of the strong energy condition
occurs. It is also interesting to analyze where this violation occurs with respect
to the characteristic scale of the energy density distribution; let us take the characteristic
scale to be at $r \sim \sqrt{2 \Theta}$. This corresponds to $x \sim 1$; we  would like
to compare this value with the maximum distance at which the strong energy condition is violated,
a consideration that can be easily made looking at figure \ref{fig:StrEneCon}, where,
together with the inner Cauchy horizon, the strong energy condition violating region
(the gray area, i.e. \emph{union} of light and dark gray areas) and the region occupied by the
characteristic width of the energy distribution (dark gray area) are emphasized.
We can, then, see that the energy violating region is of the same size (as order of magnitude)
of the characteristic width of the Gaussian energy distribution. For large enough values of the
dimensionless mass $\mu _{0}$ both regions extend beyond the inner Cauchy horizon, but none
of them ever occupies space outside the event horizon.

\subsection{\label{subsec:eneconsum}Discussion}

In this section we have seen that a recently obtained solution of Einstein equations
\cite{bib:PhLeB2006.632...547S,bib:PhLeB2007.645...261S} is, in fact, a spherically
symmetric black hole with regular center. Using the general results of subsection
\ref{subsec:genfra} this can be understood directly from the
choice of the Gaussian energy density source and from the assumptions on the stress
energy tensor: nevertheless we pedagogically derived all the standard properties of the solution.
In view of applications of this model to phenomenological situations, it is important
to study where the energy conditions are violated. From general arguments the weak
energy condition is satisfied, but the strong one must be violated. A violation of the
strong energy condition is equivalent to a violation of the attractive character of
gravity, and must be confined to scales where it can be physically explained, perhaps
using non classical effects.

We have seen that when horizons are present the violation always
takes place inside the event horizon, but the non singular character of the
metric and the fact that it is not globally  hyperbolic make this result, by itself,
not very significative: if an observer \emph{could} travel (we ignore in the following
qualitative description stability issues) from one asymptotic region I to another
(see figure \ref{fig:pendiamuGmuc}) crossing regions II, III and II again, in this
order, then (s)he might directly experience a violation of the strong energy condition.
In this respect, situations in which $\mu _{0} > \mu ^{\ast}$, shown in figure
\ref{fig:StrEneConVio002} panel c3) could present bigger troubles from the point of
view of the phenomenological interpretation, whereas cases like the one in panel c1)
in the same figure appear more sound: it thus seems that, quite generically, natural
upper bounds for $\mu _{0}$ can be obtained. Given that
$\mu _{0} = m _{0} / \sqrt{2 \Theta}$, we then see that pushing $\sqrt{2\Theta}$ to lower and
lower values requires the total mass to be also smaller and smaller, \emph{if we insist that
the strong energy condition violation has to take place inside the Cauchy horizon}.
At the same time, we point out that pushing too far away the scale at which the strong energy
condition is violated, would provide us with a solution which could be hardly distinguishable
from other ones: this suggest that for phenomenological applications it may be important to carefully
discuss the various scales present in the problem keeping in mind, both, the global causal
structure as well as the strong energy condition violating region. After this note of caution,
we think that it is interesting to observe how the violation of the energy condition
\emph{does} take place on scales comparable to the scales which characterize the
matter-energy distribution; this fact is then suggestive, since, if non classical
effects are advocated as motivations for the matter-energy source, then we see that they
actually modify the spacetime structure mostly at the scale at which they become relevant:
this is, in our opinion, an important consistency check \emph{when dealing with non linear
situations}.

Although the configurations \emph{with horizon} are the most studied and emphasized ones,
we think that the G-lump solutions are also interesting and, perhaps, closer in spirit to
what we would intuitively think as a regular spacetime, since they do not have horizons
or other structural/apparent pathologies. Notice that G-lump solutions are realized for
$\mu _{0} < \mu _{0} ^{\mathrm{cr}}$, which, after fixing the scale $\sqrt{2 \Theta}$,
can be traduced into an upper bound for their total mass-energy $m _{0}$, consistently
with their localized structure (see also the discussion in \cite{bib:IJMPD2003..12..1015D}).

It is very likely that similar considerations will hold also for other spherically symmetric
black holes with a regular center: it then becomes interesting to discuss if/how different
realizations could be practically distinguished, but we will not discuss this point here.

\section{\label{sec:discon}Conclusions}

Summarizing, we have reviewed various spherically symmetric black hole solutions with a regular
center: in the first part of this contribution we kept an historical perspective, and we tried
to emphasize alternative approaches and ideas, showing at the end that they can be all understood
inside the same general framework. Then, in the second part, we specialized to a very specific solution
and we showed how it also fits the above mentioned general framework. We analyzed in detail the
violation of the strong energy condition: results similar to the one that we have shown ere are likely to
hold for different specific realizations. If we, thus, interpret the non standard behavior of these
solutions at small scales invoking some sort of non-classical effects, it is remarkable that, self
consistently, these proposals violate known properties of matter and energy only at scales at which
these effects are supposed to be dominant. This might be \emph{not granted} in a \emph{non linear} theory
and, wether we consider the G-lump or the black hole structures,
it would be interesting to obtain these solutions in a consistent, fully non-classical treatment
of spacetime and its matter fields content, or, preliminarily, at least incorporate, effectively,
this effects also in the left hand side of Einstein equations.

\section*{Acknowledgements}

I would like to thank prof. W. Israel (for exchange on the subject and for
reading an early draft of the paper), prof. K. Bronnikov, prof. I.
Dymnikova and prof. R. Giamb\`{o} (for pointing out several related references)
and prof. V. Frolov (for discussing various aspects of singularity avoidance
in black hole spacetimes).\\
I would also like to thank the Yukawa Institute for Theoretical Physics at
Kyoto University: discussions during the KIAS-YITP joint workshop
YITP-W-07-10 on ``String Phenomenology and Cosmology'' were useful to
complete this work.\\
This work is supported by a long term invitation fellowship of
the Japan Society for the Promotion of Science (JSPS).

\end{document}